# Visualization of Gaussian Mode Profile in Gigahertz Surface-Acoustic-Wave Resonators

Shizai Chu[1], Suraj Thapa Magar[2], John Nichol[2], Keji Lai[1]

[1] Department of Physics, University of Texas at Austin, Austin TX 78712, USA

[2] Department of Physics and Astronomy, University of Rochester, Rochester NY 14627, USA

## Abstract

Surface-acoustic-wave (SAW) resonators operating at gigahertz (GHz) frequencies are widely used in wireless telecommunication and quantum information processing. Successful implementation of such resonators calls for detailed microscopic understanding of their mode profiles, energy dissipation channels, and imperfections from microfabrication. In this work, we report on the visualization of acoustic waves in LiNbO$_3$ SAW resonators by transmission-mode microwave impedance microscopy (TMIM). The Gaussian mode profile tightly confined by reflecting mirrors is vividly seen in the TMIM images, from which the linewidth of the resonator itself can be extracted. The spatially resolved acoustic profile also allows us to perform failure analysis on faulty devices. Our work establishes a pathway for further optimization of SAW resonators for classical and quantum acoustic applications.

Surface acoustic waves (SAWs), elastic excitations on the surface of solids, are increasingly being explored as low-loss on-chip carriers for advanced classical and quantum information technology [1-3]. SAW devices operating at gigahertz (GHz) frequencies are capable of mediating interactions between various quantum systems and transducing quantum information across different energy domains [4-12]. Of particular interest here are reflection-based SAW resonators, the mechanical equivalent of optical Fabry-Perot interferometers, which confine acoustic standing waves within resonant cavities and enhance their interaction with superconducting qubits or optical photons [13-16]. SAW resonators with compact dimensions and high quality (Q) factors thus play a central role in phonon storage, qubit-phonon coupling, and microwave-to-optical transduction for various quantum acoustic architectures [7,8,17,18].

SAW resonators are typically patterned on piezoelectric materials, using interdigital transducers (IDTs) for electromechanical transduction [1]. Their performance is highly dependent on geometric parameters such as the IDT pitch, reflector period, and cavity shape. To date, the design of SAW resonators has been heavily relying on finite-element modeling and conventional S-parameter characterization. On the other hand, mesoscopic device-specific information, such as aperture diffraction, SAW leakage, or disorder-induced scattering, would be very important for resonant microcavities designed to achieve strong coupling with nanoscale quantum systems [14-16]. In the GHz regime, spatially resolved studies become increasingly difficult with increasing operation frequency and decreasing acoustic wavelength. The rapidly evolving quantum acoustic research therefore calls for the development of high-resolution microscopy techniques and the imaging of GHz acoustic fields.

In this work, we report on the acoustic imaging of 6 GHz LiNbO$_3$ SAW resonators with curved reflecting mirrors by transmission-mode microwave impedance microscopy (TMIM) [19-21]. The technique operates by measuring the SAW-induced surface potential and demodulating the GHz signal in a phase-sensitive manner [22-24]. The Gaussian-like mode profile as a function of frequency can be readily mapped out by the TMIM, which enables us to extract the displacement fields and linewidth of the resonator itself. The spatially resolved acoustic fields inside and outside the microcavities also allow us to diagnose faulty devices with inferior properties. Our work contributes to the design and optimization of high-Q SAW resonators for classical and quantum information science applications.

The SAW resonators studied in this work are fabricated on 128° Y-cut LiNbO$_3$ piezoelectric substrates by electron-beam lithography [25]. As shown in the inset of Fig. 1a, the devices are engineered with the following features to achieve a small mode volume and a high electrical impedance, which are desirable for strong coupling to solid-state quantum systems [26]. First, the length of the IDT electrodes is very short (~ 7 μm) to minimize the mode volume and reduce the capacitance. Secondly, the electrically floating mirrors with 50 or 200 periods are curved to focus SAWs [18, 26-29] and create a Gaussian mode with a small beam waist. Finally, the gap between IDT electrodes and mirrors is eliminated to ensure a nearly constant effective wave speed along the direction of propagation. Fig. 1b shows the amplitude and phase of the reflection coefficient S$_{11}$ of a typical resonator with a linewidth of ~12 MHz, corresponding to an apparent Q-factor of ~ 500. Detailed design procedures and S-parameter characterizations of similar resonators can be found in Ref. [25]. We emphasize that, due to the small dimensions of the resonant cavity and short SAW wavelength (< 1 μm), it would be very challenging to image the acoustic displacement fields by traditional laser vibrometry, where the spatial resolution of ~ 0.5 μm is limited by optical diffraction.

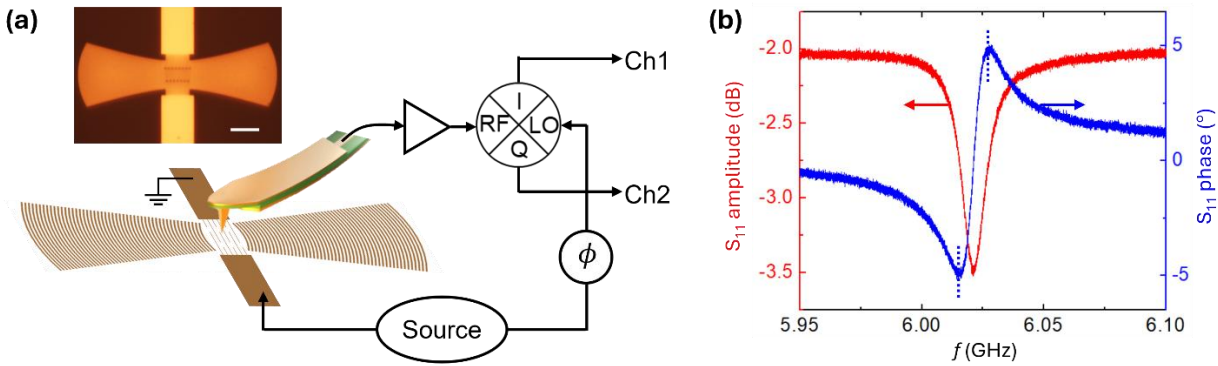

**Fig. 1. (a)** Schematic of the TMIM setup and the SAW resonator with mirror reflectors. The inset shows the optical image of a typical resonator in this study. The scale bar is 5 μm. **(b)** S$_{11}$ amplitude and phase of a normal SAW resonator. The linewidth is determined by the blue dashed lines, from which an apparent Q-factor of ~ 500 can be extracted for this resonator.

The schematic of our atomic-force microscopy (AFM) based TMIM setup is shown in Fig. 1a [19,20]. The spatial resolution of TMIM is determined by the tip diameter (~ 100 nm), which is sufficient to map out the 6 GHz SAWs in this device. The acoustic wave is launched by the IDT, and the induced surface-potential modulation is detected by the tip. The signal is then amplified and demodulated by an in-phase–quadrature (I/Q) mixer, using the same microwave source as the local oscillator (LO) reference. As shown in our previous work [22-24, 30, 31], for a propagating wave with the surface potential $V_{RF} \propto e^{i(kx-\omega t)}$, the two output signals are 90° out-of-phase between each other, e.g., $V_{Ch1} \propto \cos kx$ and $V_{Ch2} \propto \sin kx$. Fast Fourier transformation (FFT) of the complex-valued $V_{Ch1} + i * V_{Ch1}$ image can yield information on the wavevector ($k$) in reciprocal space. On the other hand, for a standing wave with $V_{RF} \propto \sin kx \cos \omega t$, one can adjust the LO phase such that the SAW features only appear in one of the two channels. The TMIM thus provides a phase-sensitive mapping of the demodulated electric potential and local displacement fields on the sample surface.

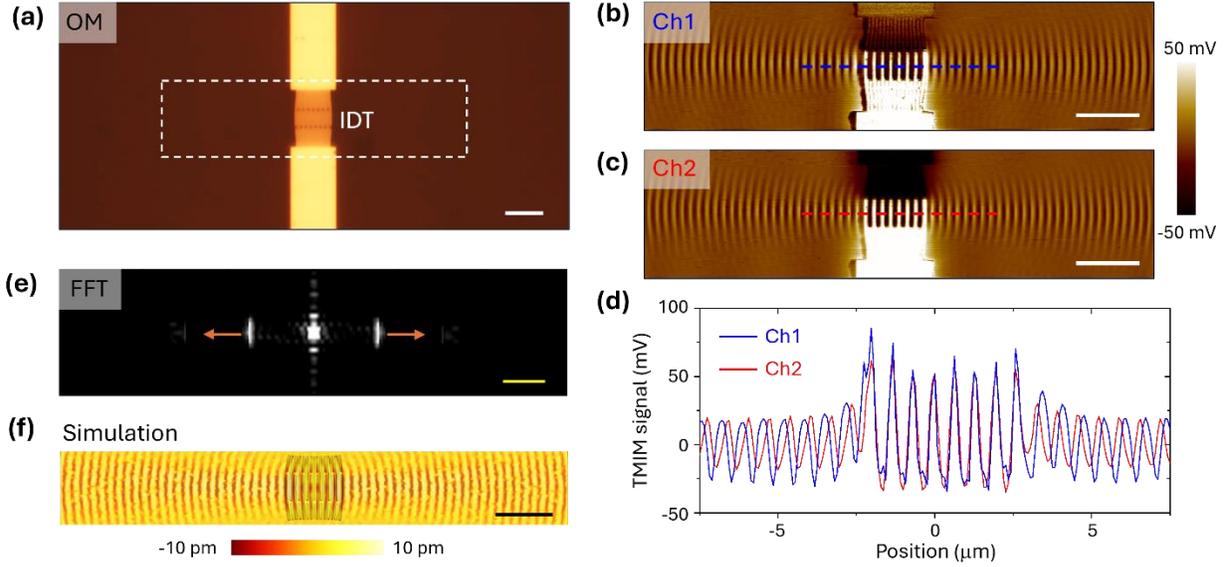

**Fig. 2. (a)** Optical microscopy (OM) image of a simple IDT device. **(b)** TMIM-Ch1 and **(c)** -Ch2 images in the dashed box of (a). **(d)** Line profiles along the dashed lines in (b) and (c). **(e)** FFT map of the TMIM images. The orange arrows represent the directions of wave propagation. **(f)** Simulated displacement fields of the propagating SAW generated by the IDT. The scale bars are 5 µm for the real-space images (a-c, f) and $2\pi \cdot \mu m^{-1}$ for the reciprocal-space image (e).

Before discussing results of the resonators, we benchmark our measurements using a device without mirror reflectors. Fig. 2a shows the optical micrograph of a bare IDT patterned on the same LiNbO₃ substrate. The corresponding TMIM images with an excitation power of −6 dBm ($V_{rms} \sim 0.2$ V for small $S_{11}$) are displayed in Figs. 2b and 2c, which clearly show the acoustic wavefronts on both sides of the IDT fingers. In Fig. 2d, we plot the line profiles of both TMIM channels across the center of the device. As expected, the signal of Ch1 away from the IDT is 90° out-of-phase from that of Ch2, indicative of propagating waves excited by the IDT. Fig. 2e shows the FFT spectral map of the combined $V_{Ch1} + i * V_{Ch1}$ data, where the two peaks correspond to outgoing waves with wavevectors $k = \pm 9.5$ µm⁻¹ or a wavelength of $\lambda = 2\pi/|k| = 0.66$ µm. Finally, we perform finite-element modeling (FEM) of the IDT structure using commercial software

COMSOL 5.6. As depicted in Fig. 2f, the simulated displacement fields on the LiNbO₃ substrate closely match the TMIM data, with an oscillation amplitude of ~ 5 pm under the same excitation power as in the imaging experiment.

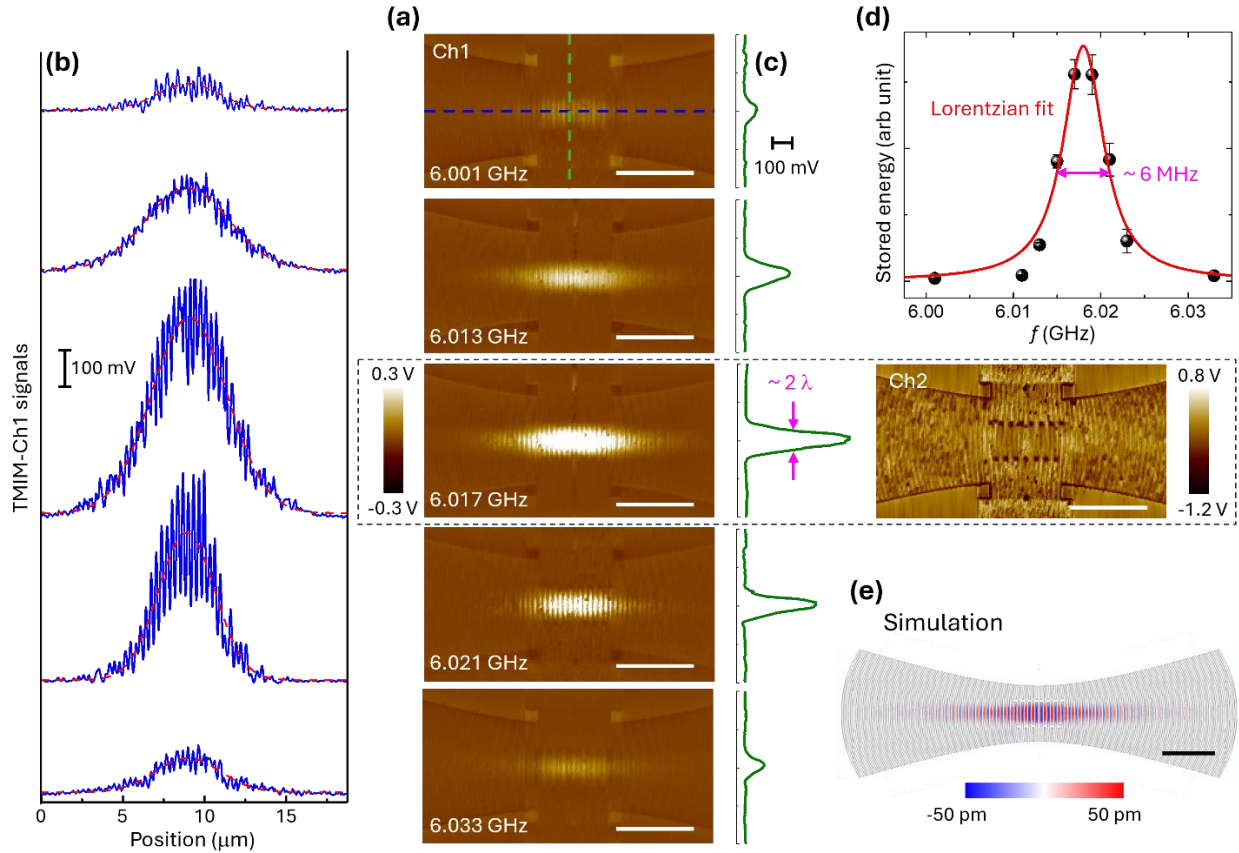

Fig. 3. (a) TMIM-Ch1 images of the SAW resonator in Fig. 1b at selected frequencies. The TMIM-Ch2 image at 6.017 GHz is also shown for comparison. **(b)** and **(c)** TMIM-Ch1 signals along the blue and green dashed lines in the 6.001 GHz image in (a), respectively. The red curves in (b) are Gaussian fits to the data. The beam waist is around 2λ, as indicated in (c). **(d)** Acoustic energy confined in the resonator, which is proportional to the square of the peak TMIM signals, as a function of frequency. The red line is a Lorentzian fit to the data. (e) Simulated acoustic displacement fields inside the mirror resonator, showing the spindle-like mode profile. All scale bars are 5 μm.

With full understanding of the IDT-only device, we now move on to discuss results on the reflector-based SAW resonator in Fig. 1b. Note that we have adjusted the LO phase at each frequency such that the standing wave pattern predominantly appears in Ch1, whereas signals in Ch2 are mostly due to topographic crosstalk. The excitation power is kept at the same level of −6 dBm. The TMIM images at selected frequencies are displayed in Fig. 3a, showing a spindle-like feature with its long axis extending several microns into the curved mirrors. The line profiles perpendicular and parallel to the IDT (dashed lines in the 6.001 GHz image Fig. 3a) are plotted in Figs. 3b and 3c, respectively, which can be fitted by Gaussian functions. The beam waist at the center of the Gaussian mode profile is 1.2 ~ 1.4 μm (about 2 λ) [25]. The TMIM signals in the cavity increase rapidly toward the resonant frequency. Note that the TMIM tip is a weak perturbation and frequency-independent load to the piezoelectric potential [19]. In other words, the square of the peak TMIM signals in Fig. 3a is proportional to the acoustic energy $E$ stored in the resonator. We then make a reasonable assumption that the energy lost per cycle $\Delta E$ is also insensitive to frequency such that Q = $E / \Delta E$ can be extracted from the $f$-dependent plot of peak $V_{Ch1}^2$ values, as shown in Fig. 3d. The full-width-at-half-maximum linewidth of the Lorentzian fit is ~ 6 MHz, corresponding to a Q-factor of 1000. Note that this linewidth is lower than that extracted from the S-parameter measurement (Fig. 1b), because the two measurements are differently affected by parasitic circuit elements, such as contact resistances and shunt capacitances. In this case, the TMIM result gives a lower linewidth that is likely closer to the intrinsic linewidth of the resonator. Finally, the Gaussian mode profile can be seen in the FEM simulation (Fig. 3e). The simulated displacement fields (~ 50 pm) are one order of magnitude greater than that of the IDT-only device (Fig. 2f), consistent with the enhanced acoustic power in the resonator. It should be noted that the simulation provides the surface displacement on the piezoelectric LiNbO₃ substrate, which oscillates between positive and negative values. On the other hand, the TMIM probe moves across the electrically floating mirrors, which may be responsible for the all-positive values in Fig. 3a. Further experiments are needed to understand the polarity of TMIM signals when the tip scans over metal electrodes.

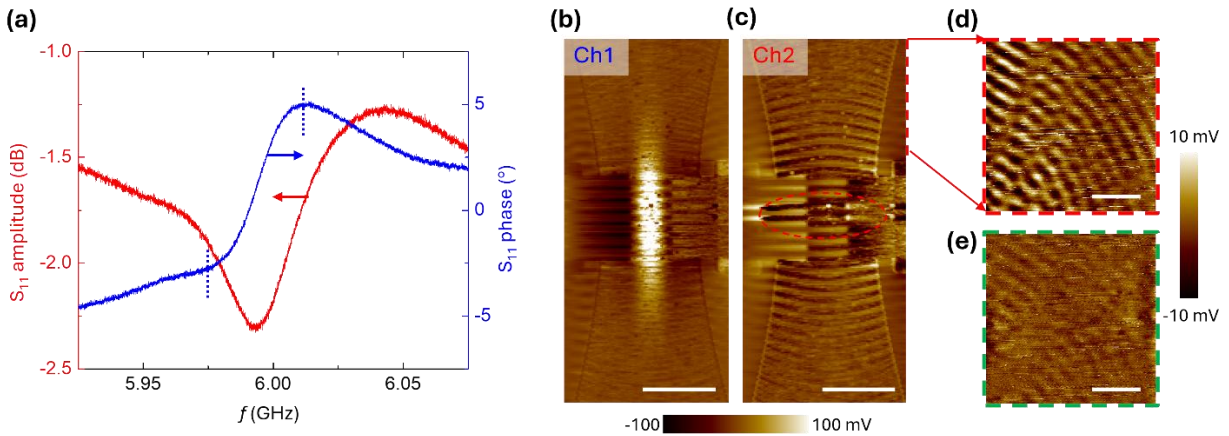

**Fig. 4. (a)** S₁₁ amplitude and phase of a faulty resonator. The linewidth is again determined by the blue dashed lines, from which a Q-factor of ~ 150 can be extracted. **(b)** TMIM-Ch1 and **(c)** -Ch2 images of the same device at its resonant frequency of 5.994 GHz. The IDT fingers in the red oval exhibit abnormal polarities, which may disrupt the electrical potential required to launch the acoustic waves.

**(d)** TMIM image outside the faulty resonator, showing substantial leakage of acoustic power. **(e)** Same as (d) for the normal resonator presented in Fig. 3, showing negligible acoustic leakage. The scale bars are 5 μm in (b, c) and 2 μm in (d, e).

The spatially resolved acoustic mapping by TMIM also allows us to perform failure analysis on faulty devices. Fig. 4a shows the $S_{11}$ spectrum of a defective resonator fabricated in the same batch as the one presented in Figs. 1 and 3. Such a terminal-to-terminal measurement can only reveal its inferior electrical property, e.g., a low Q-factor of ~ 150. The TMIM imaging, on the other hand, provides much richer local information of the device. As seen in Fig. 4b, while the spindle-like mode profile still appears in TMIM-Ch1 (adjusted to mostly contain the standing-wave pattern), the signal strength is much lower than that of the normal device in Fig. 3. Interestingly, the abnormal IDT signals in the TMIM-Ch2 image (Fig. 4c) indicate that two of the interdigital fingers may have been short-circuited during the fabrication process. In this scenario, the electrical shunt would disrupt the alternating electrical potential required to launch the targeted SAWs. Consequently, the distorted wavefronts impinge on the reflectors with incorrect phases, resulting in inefficient mode confinement and poor device performance. The conjecture is further supported by the substantial TMIM signals outside the mirror regions, as seen in Fig. 4d. In comparison, the acoustic leakage for a normal device is barely discernible in the TMIM image (Fig. 4e). We emphasize that the TMIM signal level provides a semi-quantitative understanding of the electrical characteristics. For instance, the signal outside the resonator in Fig. 4d is ~ 10% of that of the Gaussian mode in Fig. 4b. With a Q-factor of ~ 150, the result suggests that the acoustic leakage is the main loss mechanism of this faulty device. In contrast, the leakage signal in Fig. 4e is much less than 1% of the Gaussian mode signal in Fig. 3. Given that Q ~ 1000 for the working device, it is unlikely that the acoustic leakage is the dominating factor. Further improvement of device performance should thus focus on other mechanisms such as material loss and damping sources [32, 33].

In conclusion, we have successfully visualized the acoustic mode profile confined by mirror reflectors in SAW resonators. Using results on a simple IDT device as the benchmark, we achieve quantitative understanding of the TMIM signals in terms of the acoustic displacement fields. The frequency-dependent images enable us to analyze the mode profile and extract a Q-factor of the resonator, which is higher than the value determined by S-parameter measurements. The spatially resolved acoustic mapping also allows us to diagnose device problems during the fabrication process. Our work is important for designing and optimizing SAW resonators for classical and quantum acoustic applications.

## Acknowledgements

This work was primarily supported by the Gordon and Betty Moore Foundation, grants https://doi.org/10.37807/gbmf12238 and https://doi.org/10.37807/gbmf12254. The TMIM data analysis was partially supported by the National Science Foundation Division of Engineering Award ECCS-2221822 and the Welch Foundation grant F-1814.

**References:**

[1]    S. Datta, *Surface Acoustic Wave Devices* (Prentice-Hall, Englewood Cliffs, New Jersey, 1986).

[2]    D. Morgan, *Surface Acoustic Wave Filters: With Applications to Electronic Communications and Signal Processing* (Academic Press, Amsterdam, 2010).

[3]    P. Delsing, A. N. Cleland, M. J. Schuetz, J. Knörzer, G. Giedke, J. I. Cirac, K. Srinivasan, M. Wu, K. C. Balram, and C. Bäuerle, *et al.,* The 2019 surface acoustic waves roadmap, *J. Phys. D: Appl. Phys.* **52**, 353001 (2019).

[4]    M. V. Gustafsson, T. Aref, A. F. Kockum, M. K. Ekström, G. Johansson, and P. Delsing, Propagating phonons coupled to an artificial atom, *Science* **346**, 207 (2014).

[5]    M. J. A. Schuetz, E. M. Kessler, G. Giedke, L. M. K. Vandersypen, M. D. Lukin, and J. I. Cirac, Universal Quantum Transducers Based on Surface Acoustic Waves, *Phys. Rev. X* **5**, 031031 (2015).

[6]    D. A. Golter, T. Oo, M. Amezcua, K. A. Stewart, and H. Wang, Optomechanical quantum control of a nitrogen vacancy center in diamond, *Phys. Rev. Lett.* **116**, 143602 (2016).

[7]    R. Manenti, A. F. Kockum, A. Patterson, T. Behrle, J. Rahamim, G. Tancredi, F. Nori, and P. J. Leek, Circuit quantum acoustodynamics with surface acoustic waves, *Nat. Commun*. **8**, 975 (2017).

[8]    K. J. Satzinger, Y. Zhong, H.-S. Chang, G. A. Peairs, A. Bienfait, M.-H. Chou, A. Cleland, C. R. Conner, É. Dumur, and J. Grebel, *et al*., Quantum control of surface acoustic wave phonons, *Nature* **563**, 661 (2018).

[9]    A. Bienfait, K. J. Satzinger, Y. Zhong, H.-S. Chang, M.-H. Chou, C. R. Conner, É. Dumur, J. Grebel, G. A. Peairs, and R. G. Povey, *et al*., Phonon-mediated quantum state transfer and remote qubit entanglement, *Science* **364**, 368 (2019).

[10]    S. J. Whiteley, G.Wolfowicz, C. P. Anderson, A. Bourassa, H.Ma, M. Ye,G.Koolstra, K. J. Satzinger, M.V.Holt, and F. J. Heremans, et al., Spin–phonon interactions in silicon carbide addressed by Gaussian acoustics, *Nat. Phys*. **15**, 490 (2019).

[11]    S. Maity, L. Shao, S. Bogdanovic, S. Meesala, Y.-I. Sohn, N. Sinclair, B. Pingault, M. Chalupnik, C. Chia, and L. Zheng, et al., Coherent acoustic control of a single silicon vacancy spin in diamond, *Nat. Commun*. **11**, 193 (2020).

[12]    D. Awschalom, K. K. Berggren, H. Bernien, S. Bhave, L. D. Carr, P. Davids, S. E. Economou, D. Englund, A. Faraon, and M. Fejer, et al., Development of quantum interconnects (quics) for next-generation information technologies, *PRX Quantum* **2**, 017002 (2021).

[13]    J. Bochmann, A. Vainsencher, D. D. Awschalom, and A. N. Cleland, Nanomechanical coupling between microwave and optical photons, *Nat. Phys.* **9**, 712 (2013).

[14]    R. Manenti, M. Peterer, A. Nersisyan, E. Magnusson, A. Patterson, and P. Leek, Surface acoustic wave resonators in the quantum regime, *Phys. Rev. B* **93**, 041411 (2016).

[15]    B. A. Moores, L. R. Sletten, J. J. Viennot, and K. Lehnert, Cavity quantum acoustic device in the multimode strong coupling regime, *Phys. Rev. Lett.* **120**, 227701 (2018).

[16]    G. S. MacCabe, H. Ren, J. Luo, J. D. Cohen, H. Zhou, A. Sipahigil, M. Mirhosseini, and O. Painter, Nano-acoustic resonator with ultralong phonon lifetime, *Science* **370**, 840 (2020).

[17]    Y. Chu, P. Kharel, W. H. Renninger, L. D. Burkhart, L. Frunzio, P. T. Rakich, and R. J. Schoelkopf, Quantum acoustics with superconducting qubits, *Science* **358**, 199 (2017).

[18]    A. Iyer, Y. P. Kandel, W. Xu, J. M. Nichol, and W. H. Renninger, Coherent optical coupling to surface acoustic wave devices, *Nat. Commun.* **15**, 3993 (2024).

[19]    L. Zheng, D. Wu, X. Wu, and K. Lai, Visualization of surface-acoustic wave potential by transmission-mode microwave impedance microscopy. *Phys. Rev. Appl.* **9**, 061002 (2018).

[20]    L. Zheng, L. Shao, M. Loncar, and K. Lai, Imaging acoustic waves by microwave microscopy: Microwave impedance microscopy for visualizing gigahertz acoustic waves. *IEEE Microw. Mag.* **21**, 60 (2020).

[21]    J.-Y. Shan, N. Morrison, and E. Y. Ma, Circuit-level design principles for transmission-mode microwave impedance microscopy, *Appl. Phys. Lett.* **122**, 123505 (2023).

[22]    D. Lee, S. Meyer, S. Gong, R. Lu, and K. Lai, Visualization of acoustic power flow in suspended thin-film lithium niobate phononic devices, *Appl. Phys. Lett.* **119**, 214101 (2021).

[23]    D. Lee, Q. Liu, L. Zheng, X. Ma, H. Li, M. Li, and K. Lai, Direct Visualization of Gigahertz Acoustic Wave Propagation in Suspended Phononic Circuits, *Phys. Rev. Applied* **16**, 034047 (2021).

[24]    D. Lee, S. Jahanbani, J. Kramer, R. Lu, and K. Lai, Nanoscale imaging of super-high-frequency microelectromechanical resonators with femtometer sensitivity, *Nat. Commun.* **14**, 1188 (2023).

[25]    Y. P. Kandel, S. T. Magar, A. Iyer, W. H. Renninger, and J. M. Nichol, High-impedance surface-acoustic-wave resonators, *Phys. Rev. Applied* **21**, 014010 (2024).

[26]    J. Z.Wilcox and R. E. Brooks, Time-Fourier transform by a focusing array of phased surface acoustic wave transducers, *J. Appl. Phys.* **58**, 1148 (1985).

[27]    M. de Lima Jr, F. Alsina,W. Seidel, and P. Santos, Focusing of surface-acoustic-wave fields on (100) GaAs surfaces, *J. Appl. Phys.* **94**, 7848 (2003).

[28]    M. E. Msall and P. V. Santos, Focusing surface acoustic wave microcavities on GaAs, *Phys. Rev. Appl.* **13**, 014037 (2020).

[29]    R. A. DeCrescent, Z. Wang, P. Imany, R. C. Boutelle, C. A. McDonald, T. Autry, J. D. Teufel, S. W. Nam, R. P. Mirin, and K. L. Silverman, Large single-phonon optomechanical coupling between quantum dots and tightly confined surface acoustic waves in the quantum regime, *Phys. Rev. Appl.* **18**, 034067 (2022).

[30]    Q. Zhang, D. Lee, L. Zheng, X. Ma, S. I. Meyer, L. He, H. Ye, Z. Gong, B. Zhen, K. Lai, and A. T. C. Johnson, Gigahertz topological valley Hall effect in nanoelectromechanical phononic crystals, *Nat. Electron.* **5**, 157 (2022).

[31]    D. Lee, Y. Jiang, X. Zhang, S. Jahanbani, C. Wen, Q. Zhang, A.T. C. Johnson, and K. Lai, Klein tunneling of gigahertz elastic waves in nanoelectromechanical metamaterials, *Device* **2**, 100474 (2024).

[32]    G. Pillai and S.S. Li, Piezoelectric MEMS resonators: A review, *IEEE Sensors Journal* **21**, 12589 (2020).

[33]    K. Julius, J. V. Knuuttila, T. Makkonen, V. P. Plessky, and M. M. Salomaa, Acoustic loss mechanisms in leaky SAW resonators on lithium tantalate, *IEEE Trans. Ultrason. Ferroelectr. Freq. Control* 48, 1517 (2001).